\documentclass[11pt]{article}
\usepackage{amssymb,amsmath,cite,geometry,graphicx,url,mathrsfs}
\catcode`\@=11

\@addtoreset{equation}{section}
\def\theequation{\arabic{section}.\arabic{equation}}
     
\catcode`\@=11
\def\thesection{\arabic{section}}

\def\appendix{\setcounter{section}{0}
        \def\thesection{Appendix.}
        \def\theequation{\Alph{section}.\arabic{equation}}}
\def\section{\@startsection{section}{1}{\z@}{3.5ex plus 1ex minus
   .2ex}{2.3ex plus .2ex}{\large\bf}}
     
\long\def\@makefntext#1{\parindent 0cm\noindent
\hbox to 1em{\hss$^{\@thefnmark}$}#1}

\newcommand{\captionfonts}{\small}
\makeatletter  
\long\def\@makecaption#1#2{%
  \vskip\abovecaptionskip
  \sbox\@tempboxa{{\captionfonts #1: #2}}%
  \ifdim \wd\@tempboxa >\hsize
    {\captionfonts #1: #2\par}
  \else
    \hbox to\hsize{\hfil\box\@tempboxa\hfil}%
  \fi
  \vskip\belowcaptionskip}
\makeatother   

\def\topx{\mathop{\raise4pt\hbox{$\vee$}\hspace{-.7em}\lower4pt\hbox{$\wedge$}}}

\begin{document}
\begin{flushright}
December 2023\\  
  \end{flushright}
\begin{center}
{\Large\bf
 Quantum Gravity in 2+1 Dimensions}\\  
\vspace{2ex}
{S.~C{\sc arlip}\footnote{\it email: carlip@physics.ucdavis.edu}\\
       {\small\it Department of Physics, 
        University of California}\\
       {\small\it Davis, CA 95616, USA}}
       
\end{center}
\vspace*{1ex}
 
\section{Introduction}

It has been understood since the early 1960s that general relativity becomes drastically
simpler in three spacetime dimensions \cite{Staruszkiewicz}.  In three dimensions, the Weyl
tensor vanishes identically, and the full curvature tensor is determined algebraically by the 
Ricci tensor:
\begin{equation}
R_{\mu\nu}{}^{\rho\sigma} = \delta_\mu^\rho R_\nu^\sigma
   + \delta_\nu^\sigma R_\mu^\rho  - \delta_\nu^\rho R_\mu^\sigma
   - \delta_\mu^\sigma R_\nu^\rho - \frac{1}{2}
  \left(\delta_\mu^\rho\delta_\nu^\sigma - \delta_\mu^\sigma\delta_\nu^\rho\right)R .
\label{a1}
\end{equation}
Thus if the Einstein field equations are satisfied, the full curvature is determined algebraically by
the stress-energy tensor; there are no local gravitational degrees of freedom, no gravitational
waves.  In particular, if the vacuum field equations with a cosmological constant $\Lambda$ holds,
\begin{equation}
R_{\mu\nu} = 2\Lambda g_{\mu\nu} ,
\label{a2}
\end{equation}
then  spacetime is a manifold of constant curvature.  

This absence of local degrees of freedom can be confirmed by a simple counting argument.
In three dimensions, the metric $g_{\mu\nu}$ has six independent components.  Three of
these can be fixed by coordinate choices, leaving three potential physical degrees of freedom.
But three of the Einstein field equations---the equations $G^0_\mu = \kappa^2 T^0_\mu$---are
constraints, containing only first time derivatives.  These determine the three remaining
components of the metric, leaving no free initial data.   This  is reflected in the weak 
field approximation, where small perturbations of the metric are ``pure gauge,'' and can be 
eliminated by a change of coordinates \cite{Carlip_rev}.  By a similar argument, 
the theory has no Newtonian limit \cite{Carlip_book}: two point particles at rest experience 
no attraction, though particles in relative motion can deflect each other \cite{DesJacb}.

This simplicity makes (2+1)-dimensional gravity a good model for exploring 
conceptual issues of quantum gravity.  At first sight, though, the model appears  
\emph{too} simple, seemingly lacking any interesting gravitational degrees of freedom.  Indeed,
early papers  (for instance, \cite{Staruszkiewicz,DesJacb,DesJaca,DJtH,tHooft}) focused on 
point particle solutions, whose only dynamical degrees of freedom are particle locations.
It is now understood, though, that even in three dimensions, interesting gravitational degrees
of freedom can arise from two sources:
\begin{itemize}
\item In a topologically nontrivial spacetime, even with constant curvature, global geometric
degrees of freedom can be present.  These arise as holonomies around noncontractible
curves, or equivalently as ``geometric structures'' describing the gluing of local constant curvature 
patches.
\item In a spacetime with a boundary (or an asymptotic boundary), the Einstein-Hilbert action has
extrema only if appropriate boundary conditions are imposed.  Diffeomorphisms that do not respect
these conditions are no longer invariances of the theory, and configurations that differ by such
diffeomorphisms are physically distinct;  such ``would-be gauge degrees of freedom'' (or ``edge 
modes'' or ``boundary gravitons'') become dynamical.
\end{itemize}

Thanks to this combination of comparative simplicity, physical interest, and connections to a  
variety of intriguing mathematical  structures, (2+1)-dimensional gravity has been an extremely 
active area of research for the past 25 years.  This article will give a general, though necessarily
quite incomplete, overview.

\section{Classical gravity}

We begin with classical gravity in 2+1 dimensions, with a possible cosmological constant but no other
matter.  The Einstein-Hilbert action depends on Newton's constant $G$, and while this does not
affect vacuum solutions, it appears in the symplectic structure, and therefore in quantization.
For simplicity, denote $\kappa^2 = 8\pi G$.  The action then takes the form
\begin{equation}
I_{\hbox{\scriptsize grav}} = \frac{1}{2\kappa^2}\int_M\!d^3x\sqrt{|g|}\left(R-2\Lambda\right)  .
\label{aa1}
\end{equation}
The  resulting field equations are (\ref{a2}), so solutions are spacetimes of constant curvature.  These 
can be characterized in different ways, which suggest different approaches to quantization.

\subsection{Geometric structures \label{GS}}

We first observe that any solution of the vacuum field equations (\ref{a2}) in 2+1 dimensions can 
be obtained by isometrically gluing pieces of a fixed constant curvature ``model space.'' Let $M$ be a 
three-dimensional Lorentzian manifold of constant curvature $\Lambda$.  Any point in $M$ is contained
 in a contractible neighborhood isometric to a constant curvature model space $X_\Lambda$, where 
 $X_\Lambda$ is de Sitter space if $\Lambda>0$, Minkowski space if $\Lambda=0$, or anti-de Sitter 
 space if $\Lambda<0$.  Denote the isometry group of $M$ as $G_\Lambda$, where
\begin{equation}
G_\Lambda =  
\begin{cases}
\, \hbox{SO}(3,1) &\hbox{if  $\Lambda>0$},\\
\, \hbox{ISO}(2,1) &\hbox{if  $\Lambda=0$},\\
\, \hbox{SO}(2,2) &\hbox{if  $\Lambda<0$} .
\end{cases}
\end{equation}

\begin{figure}
\begin{center}
\scalebox{.55}{
\begin{picture}(100,140)(0,-35)
\thicklines
\put (38,13){\oval(40,45)}
\put (42,40){\oval(40,45)}
\put (37,66){\oval(40,45)}
\put (32,-16){\oval(40,45)}
\put (33,91){\oval(40,45)}
\put (52,66){\vector(3,1){74}}
\put (80,82){$\phi_2$}
\put (55,38){\vector(1,0){80}}
\put (92,43){$\phi_1$}
\put (52,14){\vector(4,-1){66}}
\put (84,10){$\phi_n$}
\put (49,96){\vector(1,1){22}}
\put (50,111){$\phi_3$}
\put (123,85){\framebox(40,40)[br]{\raisebox{1ex}{$X_2\,$}}}
\put (130,20){\framebox(40,40)[br]{\raisebox{1ex}{$X_1\,$}}}
\put (115,-35){\framebox(40,40)[br]{\raisebox{1ex}{$X_n\,$}}}
\put (67,114){\framebox(40,40)[br]{\raisebox{1ex}{$X_3\,$}}}
\put (3,13){$U_n$}
\put (9,38){$U_1$}
\put (3,62){$U_2$}
\put (-3,87){$U_3$}
\put (-14,-16){$U_{n-1}$}
\put (140,87){\vector(1,-3){10}}
\put (151,71){$g_1$}
\put (150,22){\vector(-1,-2){10}}
\put (150,10){$g_n$}
\put (105,128){\vector(2,-1){20}}
\put (112,128){$g_2$}
\linethickness{1.5pt}
\qbezier(18,-60)(50,55)(20,140)
\qbezier(20,140)(0,180)(-29,130)
\qbezier(-29,130)(-54,60)(-29,-40)
\qbezier(-29,-40)(-5,-99)(18,-60)
\put(10,134){$\boldsymbol{\gamma}$}
 \end{picture}
 }
\end{center}
\caption{\small A curve $\gamma$ and a portion of its holonomy \label{fig1}}
\end{figure}
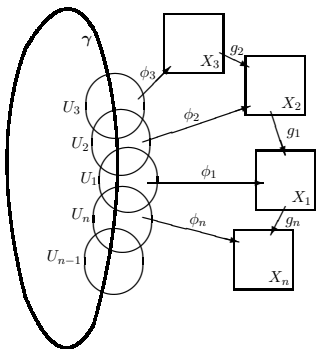

Now let $\{U_i\}$ be a collection of neighborhoods isometric to $G_\Lambda$ that cover $M$, with  isometric 
 ``coordinate maps'' $\{\phi_i\colon U_i\rightarrow X_\Lambda\}$, where the coordinates are taken 
 to be points in $X_\Lambda$ and the transition functions $g_{ij}=\phi_i\circ\phi_j{}^{-1}|_{U_i\cap U_j}$ 
 are elements of  $G_\Lambda$.  
Such a construction is an example of what what Thurston calls a geometric  structure, or a 
$(G_\Lambda,X_\Lambda)$ manifold \cite{Thurston,Canary}.  Geometric structures are characterized
by the holonomies of noncontractible curves.  Let $\gamma$ be a closed curve in $M$, covered by
coordinate patches $U_i$ with transition functions $g_i\in G_\Lambda$ such that
\begin{align}
\phi_i|_{U_i\cap U_{i+1}} &= g_i\circ \phi_{i+1}|_{U_i\cap U_{i+1}}\nonumber\\
\phi_n|_{U_n\cap U_{1}} &= g_n\circ \phi_{1}|_{U_n\cap U_{1}} .
\label{ab1}
\end{align}
The holonomy $H[\gamma] = g_1\circ\cdots\circ g_n$ then measures the failure of the transition 
functions to match up around $\gamma$, that is, the obstruction to covering $\gamma$ by a single
patch (see figure \ref{fig1}) .

It may be shown that the holonomy of a curve $\gamma$ depends only on its homotopy class 
\cite{Thurston}.  Given a fixed base point, the holonomies form a group homomorphic to $\pi_1(M)$,
$$H\colon \pi_1(M)\rightarrow G_\Lambda.$$  
There remains a bit of freedom, though: we can conjugate all of the transition functions $g_i$ 
by a fixed $h\in G_\Lambda$ without changing the geometric structure.  The space of holonomies
is thus a quotient
\begin{align}
&\mathcal{M} = {\mathop{Hom}}_0(\pi_1(M), G_\Lambda)/\sim  \nonumber\\
& \rho_1\sim\rho_2 \ \hbox{if $\rho_2 = h\cdot\rho_1\cdot h^{-1}$, $h\in G_\Lambda$} .
\label{ab2}
\end{align}
Note that not every homomorphism from $\pi_1(M)$ to $G_\Lambda$ corresponds to a smooth 
manifold.  The full space of homomorphisms  is not connected, and in general only one connected 
component yields smooth geometries \cite{Goldman}; the subscript $0$ in (\ref{ab2}) denotes the 
restriction to this component.  Although the holonomies are nonlocal, as quantum gravitational 
observables must be \cite{Torre,Donnelly}, there are algorithms for obtaining them from 
measurements of light rays \cite{Meusa}. 

A crucial question is how uniquely such a holonomy group determines the 
spacetime $M$.  If $\Lambda\le 0$ and $M$ has the topology $\mathbb{R}\times\Sigma$ with 
compact spatial slices $\Sigma$, Mess has shown that the holonomy group $H$ determines
a unique maximal domain of dependence \cite{Mess}, and that this domain of dependence 
can be expressed as a quotient $M = X_0/H, X_0\subset X_\Lambda$ \cite{Mess,Messb}.  For
$\Lambda>0$ a similar result holds \cite{Scannell}, though the situation is slightly more complicated, 
essentially because the existence of Cauchy horizons means that some locally de Sitter spacetimes
$\mathbb{R}\times\Sigma$ are not domains of dependence.

For topologies  $M= \mathbb{R}\times\Sigma$, we have $\pi_1(M) = \pi_1(\Sigma)$.  If $\Sigma$ is 
orientable and has genus $g$, the space of holonomies (\ref{ab2}) has dimension $12g-12$ for 
$g>1$, $4$ for $g=1$, and $0$ for $g=0$, that is, twice the dimension of the Teichm{\"u}ller space 
of $\Sigma$.  This is not accidental.  The Teichm{\"u}ller space ${\cal T}(\Sigma)$  has a description 
similar to (\ref{ab2}), but with $G_\Lambda$ replaced with \hbox{PSL}(2,$\mathbb{R}$) \cite{Goldmanb}.  
This permits a description of the space of vacuum spacetimes  $\mathbb{R}\times\Sigma$ 
in terms of ${\cal T}(\Sigma)$ \cite{Mess,Scannell}:
\begin{equation}
\mathcal{M}\left|_{\mathbb{R}\times\Sigma}\right. =
\begin{cases}
\,  {\cal CP}(\Sigma) &\hbox{if  $\Lambda>0$},\\
\,  T^*{\cal T}(\Sigma) &\hbox{if  $\Lambda=0$},\\
\, {\cal T}(\Sigma)\times{\cal T}(\Sigma) &\hbox{if  $\Lambda<0$} ,
\end{cases}
\label{ab3}
\end{equation}
where ${\cal CP}(\Sigma)$ is the space of complex projective structures.  This characterization 
allows a fairly explicit construction of the spacetimes in terms of domains of dependence
 \cite{Benedetti,Benedettib}, which in turn are described by a generalization of Thurston's method 
of earthquakes and grafting along measured laminations \cite{Thurstonb}.  Using the symplectic
structure described in \S\ref{CS}, one can show that these earthquake and grafting 
transformations, which parametrize motion through $\mathcal{M}$, are generated by the holonomies
in a way that can be described quite explicitly \cite{Meusx,Meusy}.   

Along with a set of solutions, classical general relativity gives us a symplectic structure on the
space of solutions, which largely determines the commutators in the quantum theory.  This symplectic 
structure can be obtained from the canonical form of the theory, but also from the boundary terms in 
the variation of the action \cite{Bombelli,Crn,Crnb,Lee}.  The fundamental group $\pi_1(\Sigma)$ of 
a surface also has a symplectic structure \cite{Goldmanb,GoldHam}, and we shall see below that 
the two are closely related.

\subsection{Chern-Simons formalism \label{CS}}

An alternative approach to classical (2+1)-dimensional gravity starts with the first order 
formalism.  The fundamental variables are a (co)frame (or triad or dreibein) one-form and
a spin connection,
\begin{equation}
e^a = e^a{}_\mu dx^\mu, \quad \omega^a = \frac{1}{2}\epsilon^{abc}\omega_{\mu bc}dx^\mu,
\label{ac1}
\end{equation}
where the metric is $g_{\mu\nu}=\eta_{ab}e^a{}_\mu e^b{}_\nu$ and, as in the Palatini 
formalism, the frame and spin connection are treated as independent variables.  The
Einstein-Hilbert action becomes
\begin{equation}
I = \frac{1}{\kappa^2}\int_M\!\left\{ e^a\wedge%
   \left(d\omega_a + \frac{1}{2}\epsilon_{abc}\omega^b\wedge\omega^c\right)
   - \frac{\Lambda}{6}\epsilon_{abc}e^a\wedge e^b\wedge e^c\right\} ,
\label{ac1a}
\end{equation}
and the vacuum Einstein field equations take the form
\begin{align}
T_a &= de_a + \epsilon_{abc}\omega^b\wedge e^c = 0 ,
\label{ac2}\\
R_a &= d\omega_a + \frac{1}{2}\epsilon_{abc}\omega^b\wedge\omega^c 
  = \frac{\Lambda}{2}\epsilon_{abc} e^b\wedge e^c .
\label{ac3}
\end{align}
Eqn.\ (\ref{ac2}) is a condition of vanishing torsion, and if the triad is invertible, it gives
the standard expression for the spin connection in terms of $e^a{}_\mu$.  Eqn.\ (\ref{ac3})
is then the vacuum Einstein equations.  Note that while this formulation is almost 
equivalent to the standard metric version of the field equations, it is not quite identical:
it allows additional solutions in which the triad is noninvertible, and the sign of $e^a{}_\mu$
is undetermined.

As Ach{\'u}carro and Townsend observed \cite{Achucarro} for $\Lambda<0$, and Witten  
expanded upon \cite{Wittena,Wittenb} for arbitrary $\Lambda$, the first order action
may be reinterpreted as a Chern-Simons action for a gauge field 
$A = e^a\mathcal{P}_a + \omega^a\mathcal{J}_a$, where
\begin{equation}
[\mathcal{J}_a,\mathcal{J}_b]=\epsilon_{ab}{}^c\mathcal{J}_c,\quad 
[\mathcal{J}_a,\mathcal{P}_b]=\epsilon_{ab}{}^c\mathcal{P}_c,\quad 
[\mathcal{P}_a,\mathcal{P}_b]=-\Lambda\epsilon_{ab}{}^c\mathcal{J}_c 
\label{ac3a}
\end{equation}
are the generators of the Lie algebra of $G_\Lambda$ (or a covering group).  For
$\Lambda\ne0$, the commutators (\ref{ac3a}) can be put in a more standard form by writing
\begin{equation}
\mathcal{J}_a^\pm  = \frac{1}{2}\left(\mathcal{J}_a\pm\frac{1}{\sqrt{\Lambda}}\mathcal{P}_a\right),
\quad A = A^{(+)a}\mathcal{J}^+_a + A^{(-)a}\mathcal{J}^-_a  ,
\label{ac3b}
\end{equation}
giving generators that satisfy the standard $\hbox{SO}(3,1)$ or $\hbox{SO}(2,2)$ commutation 
relations.  The Chern-Simons action
 \begin{equation}
I_{\mathit{CS}}[A] = \frac{1}{2\kappa^2} \int_M \mathrm{Tr}
  \left\{ A\wedge dA + \frac{2}{3} A\wedge A\wedge A \right\} 
\label{ac4}
\end{equation}
is then equivalent to (\ref{ac3}), where $\mathrm{Tr}(\mathcal{P}_a\mathcal{J}_b) = \eta_{ab}$
defines an invariant bilinear form on the Lie algebra (\ref{ac3a}).  

The Chern-Simons field equations require that the field strength of $A$---in fiber bundle terms, 
the curvature of the connection $A$---must vanish.  Such a flat connection is completely determined 
by its Wilson loops, or holonomies,
\begin{equation}
U_\gamma = P\exp\left\{ -\int_\gamma A \right\}  ,
\label{ac5}
\end{equation}
where $\gamma$ is a noncontractible closed curve and $P$ denotes path ordering.  
While the term ``holonomy'' has a slightly different meaning here than it does  for a
geometric structure,  the two are actually equivalent.  A $(G,X)$ structure
on a manifold $M$ determines a flat $G$ bundle, obtained by taking the product bundle 
$G\times U_i$ on each ``coordinate patch'' $U_i$  and using the
transition functions $g_{ij}$ of the geometric structure to glue the fibers on the overlaps
\cite{Goldmanc}.  The moduli space of flat connections is again given by (\ref{ab2}),
with the same restrictions needed to obtain a smooth geometry.

It is now straightforward to calculate the symplectic structure on the space of solutions.
The only term involving derivatives in the first order Einstein-Hilbert action (\ref{ac1a}) is
of the form $e^a\wedge d\omega_a$, so the standard covariant canonical construction 
\cite{Lee} gives a symplectic form
\begin{equation}
\Omega[(\delta e,\delta\omega)_{(1)},(\delta e,\delta\omega)_{(2)}]
   = \frac{1}{\kappa^2}\int_\Sigma \left(\delta e^a_{(1)}\wedge\delta\omega_{a(2)}
   - \delta e^a_{(2)}\wedge\delta\omega_{a(1)}\right)
\label{ac6}
\end{equation}
on the space of solutions, or equal time Poisson brackets 
\begin{equation}
\{ e^a{}_i(x), \omega_{bj}(x') \}
  = \kappa^2 \delta^a_b \epsilon_{ij}\delta^2(x-x') .
\label{ac7}
\end{equation}
These brackets extend to the holonomies (\ref{ac5}).  Because of the delta function in
(\ref{ac7}), two holonomies $U_\gamma$ and $U_{\gamma'}$ will have nonvanishing brackets 
only if the curves $\gamma$ and $\gamma'$ intersect.  At an intersection, the brackets will be
closely related to those of Goldman's symplectic structure on $\pi_1(M)$ \cite{Goldmanb,GoldHam},
but altered by the combination of $e$ and $\omega$ in (\ref{ac7}).  For $\Lambda=0$, this
reflects the cotangent bundle structure $T^*{\cal T}(\Sigma)$ of the space of solutions, while
for $\Lambda\ne0$ it implies that the holonomies of the connections $A^{(+)}$ and $A^{(-)}$ 
in (\ref{ac3b}) are independent, with nonvanishing brackets only within each sector. 

\subsection{ADM formalism \label{ADM}}

The ADM formalism offers a very different approach to classical gravity in 2+1 dimensions.  We 
start with the usual Arnowitt-Deser-Misner splitting of the metric into ``space'' and ``time'' \cite{ADM},
\begin{equation}
ds^2 = N^2dt^2 - q_{ij}(dx^i + N^i dt)(dx^j + N^j dt) ,
\label{ad1}
\end{equation}
and the usual Hamiltonian form of the action,
\begin{equation}
I_{\hbox{\scriptsize grav}} 
  = \int dt\int\nolimits_\Sigma d^2x \bigl(\pi^{ij}{\dot q}_{ij}
               - N^i{\cal H}_i -N{\cal H}\bigr) .
\label{ad2}
\end{equation}
Here $\Sigma$ is a slice of constant time with extrinsic curvature $K_{ij}$,  $q_{ij}$ is the 
induced spatial metric,  $\pi^{ij} = \frac{1}{2\kappa^2 }\sqrt{q}\,(K^{ij}- q^{ij}K)$ is the canonical 
momentum conjugate to $q_{ij}$, and the momentum and Hamiltonian constraints are
\begin{equation}
{\cal H}_i = -2{}^{\scriptscriptstyle(2)}\nabla_j\pi^j{}_ i,  \qquad
{\cal H} = \frac{2\kappa^2}{\sqrt{q}} q_{ij}q_{kl}(\pi^{ik}\pi^{jl}-\pi^{ij}\pi^{kl})
                 - \frac{1}{2\kappa^2}\sqrt{q}({}^{\scriptscriptstyle(2)}\!R - 2\Lambda) .
\label{ad3}
\end{equation}

The strategy pioneered by Moncrief in this context \cite{Moncrief} is the reduced phase 
space approach: first solve the constraints, then insert the solutions back into the
action to obtain the dynamics of the physical degrees of freedom.  One can use the York time
slicing \cite{York} $T=-K= \kappa^2 q_{ij}\pi^{ij}/\sqrt{q}$; while this choice is not always possible
in four spacetime dimensions, it is a good global choice for (2+1)-dimensional spacetimes
with topology $\mathbb{R}\times\Sigma$ \cite{Andersson,Barbot}.  The spatial metric $q_{ij}$
can always be written as \cite{Abikoff}
\begin{equation}
q_{ij} = e^{2\lambda}{\bar q}_{ij}(m_\alpha) ,
\label{ad4}
\end{equation}
where the $\bar q_{ij}(m_\alpha)$ are a finite-dimensional family of  metrics on $\Sigma$ with 
constant curvature $k=1$ for $\Sigma=S^2$, $k=0$ for $\Sigma=T^2$, and $k=-1$ for higher
genus surfaces.  Such metrics  are parametrized by a finite set of moduli $m_\alpha$, which  
constitute the Teichm{\"u}ller space $\mathcal{T}(\Sigma)$.
 
The details of the corresponding decomposition of the conjugate momentum may be found 
in \cite{Moncrief}.  The trace of $\pi^{ij}$ is essentially the York time, while the trace-free part 
can be written as a holomorphic quadratic differential, a cotangent vector on Teichm{\"u}ller 
space \cite{Abikoff}.  As in the geometric structures approach (\ref{ab3}), although from a
very different starting point, the reduced phase space is the cotangent bundle of 
Teichm{\"u}ller space.

With this decomposition, the momentum constraints are trivial, while the Hamiltonian 
constraint  becomes an elliptic differential equation for the scale factor $\lambda$,
\begin{equation}
\bar\Delta\lambda - \frac{1}{4}(T^2-4\Lambda)e^{2\lambda}  
  +  \frac{1}{2}\left[  {\bar q}^{-1}{\bar q_{ij}(m_\alpha)\bar q_{kl}(m_\alpha)
  p^{ik}(p^\alpha)p^{jl}(p^\alpha)}\right] e^{-2\lambda} - \frac{k}{2} = 0 ,
\label{ad5}
\end{equation}
where 
\begin{equation}
p^\alpha = \int_\Sigma d^2x\,
  p^{ij}\frac{\partial\ \ }{\partial m_\alpha}{\bar q}_{ij} 
\label{ad6}
\end{equation}
are the momenta conjugate to $m_\alpha$.  The theory of elliptic equations guarantee that 
(\ref{ad6}) has a unique solution, although for genus $g>1$ it cannot be written in closed form.  
The canonical action (\ref{ad2}) simplifies to a reduced phase space action 
\begin{equation}
I_{\hbox{\scriptsize grav}}
  = \int dT \left( p^\alpha \frac{dm_\alpha}{dT} - H(m,p,T) \right)  \quad \hbox{with\ 
  $\displaystyle H = \int_{\Sigma_T} d^2x \sqrt{\bar q}\, e^{2\lambda(m,p,T)}$}.
\label{ad7}
\end{equation}
We can read off the reduced phase space Poisson brackets,
\begin{equation}
\{m_\alpha, p^\beta\} = \delta_\alpha^\beta , \quad
\{m_\alpha, m_\beta\} = \{p^\alpha, p^\beta\} = 0 ,
\label{ad8}
\end{equation}
again corresponding, at least locally, to a cotangent bundle over Teichm{\"u}ller space.

\subsection{Large diffeomorphisms \label{large}}

The discussion so far has omitted an important detail, the existence of ``large'' diffeomorphisms,
diffeomorphisms that cannot be continuously deformed to the identity.  For a two-dimensional 
surface, the archetype of such a diffeomorphism is a Dehn twist, a transformation in which the
surface is cut along a noncontractible curve and then reglued with a $2\pi$ twist.  The group 
$\mathcal{D}(M)$ of such diffeomorphisms---also known as the mapping class group---acts 
on $\pi_1(M)$ as a group of outer automorphisms; for a surface $\Sigma$,  it \emph{is} the group of
outer automorphisms of $\pi_1(\Sigma)$ \cite{Harvey}.  $\mathcal{D}(M)$ thus also act on the 
space $\mathcal{M}$ of vacuum spacetimes, as well as on the spatial metric $q_{ij}$ in the
ADM formalism.  For a spacetime in which $\Sigma$ is a surface with a single puncture, the  
action of $\mathcal{D}(M)$ on the puncture actually determines the geometric structure
\cite{Carlip_meas}; this is the way a point particle ``sees'' the holonomy.
 
Large diffeomorphisms are invariances of the Einstein-Hilbert action, but they are not
generated by the constraints (\ref{ad3}).  The standard, although not universal, lore is that
they should therefore be treated as symmetries rather than invariances of a quantum theory.
 
\subsection{The spatially open case \label{open}}

The focus so far has been on spacetimes $\mathbb{R}\times\Sigma$ with $\Sigma$ closed.
Some of the work, for instance \cite{Benedetti}, allows for the introduction of punctures, and 
for $\Lambda<0$ there has been an effort to find extensions to spatially open ``multi-black hole'' 
manifolds \cite{Bonsante,Scarinci}.  In general, though, the  characterization of spatially open
vacuum spacetimes  is much more poorly understood.

The open case does, however, introduce an important new issue,  the 
need for boundary conditions and the nature of boundary symmetries.  For $\Lambda=-1/\ell^2<0$,
the natural boundary conditions for a spatially open spacetime $\mathbb{R}\times\Sigma$
are asymptotically anti-de Sitter ones.  Exact anti-de Sitter space,
\begin{equation}
ds^2 \sim -\left(1+\frac{r^2}{\ell^2}\right)dt^2 
     + \left(1+\frac{r^2}{\ell^2}\right)^{-1}dr^2 +  r^2d\varphi^2 ,
\label{ae1}
\end{equation}
has an isometry group $\hbox{SO}(2,2)\approx\hbox{SL}(2,\mathbb{R})\otimes\hbox{SL}(2,\mathbb{R})$.
As Deser and Jackiw noted \cite{DesJaca,DesJacb}, though, (2+1)-dimensional gravity is generically 
asymptotically conical.  A suitable set of boundary conditions that allow such behavior is \cite{Brown}
\begin{equation}
g_{\mu\nu} \sim \left( \begin{array}{ccc}
    -\frac{r^2}{\ell^2} + \mathcal{O}(1) & \mathcal{O}\left(\frac{1}{r^3}\right) & \mathcal{O}(1)\\[1ex]
    &\frac{\ell^2}{r^2} + \mathcal{O}\left(\frac{1}{r^4}\right) & \mathcal{O}\left(\frac{1}{r^3}\right)\\[1ex]
    && r^2 + \mathcal{O}(1) \end{array} \right) .
\label{ae2}
\end{equation}

As Brown and Henneaux showed \cite{Brown}, the group of asymptotic symmetries that preserve
these boundary conditions is much larger that the AdS isometry group; it is the two-dimensional
conformal group $\mathit{Diff}S^1\times\mathit{Diff}S^1$, generated by diffeomorphisms
$\xi(t+\varphi/\ell)$ and ${\bar\xi}(t-\varphi/\ell)$ (see \cite{Strominger,Carlip_conf} for an
explicit representation).  The generators $(L[\xi],{\bar L}[{\bar\xi}])$ of the asymptotic symmetries
satisfy a Virasoro algebra,
\begin{align}
\{L[\xi],L[\eta]\} &= L[\xi\eta'-\eta\xi'] 
   + \frac{c}{48\pi}\int\!d\varphi\,(\xi'\eta^{\prime\prime} - \eta'\xi^{\prime\prime})  \nonumber\\
\{L[\xi],{\bar L}[{\bar\eta}]\} &= 0
\label{ae3}\\
\{{\bar L}[{\bar\xi}],{\bar L}[{\bar\eta}]\} &= {\bar L}[{\bar \xi}{\bar \eta}'-{\bar\eta}{\bar\xi}'] 
   + \frac{{\bar c}}{48\pi}\int\!d\varphi\,({\bar\xi}'{\bar\eta}^{\prime\prime} - {\bar\eta}'{\bar\xi}^{\prime\prime})
\nonumber
\end{align}
with central charges
\begin{equation}
c={\bar c} = \frac{3\ell}{2G} .
\label{ae4}
\end{equation}
In hindsight, the appearance of the conformal algebra is not so surprising; the conformal boundary of
an asymptotically AdS${}_3$ spacetime is a two-dimensional cylinder, which has the conformal group
as its symmetry group.  This was perhaps the first hint of an AdS/CFT correspondence.  The appearance
of a central term is more unexpected.  It arises because in the presence of  an asymptotic boundary,
the Hamiltonian and momentum constraints (\ref{ad3})---which generate the diffeomorphisms---acquire
boundary terms.  Such terms can alter the Poisson algebra \cite{Brown,Carlip_conf}, in particular
inducing a central term.

Given the asymptotic behavior (\ref{ae2}), a theorem of Fefferman and Graham guarantees the
existence of a formal asymptotic expansion of the metric that solves the vacuum field equations
\cite{Fefferman}.  There exist coordinates for which
\begin{equation}
ds^2 = \frac{\ell^2}{r^2}dr^2 + \frac{r^2}{\ell^2}g_{ij}(r,x)dx^idx^j \quad 
   \hbox{with $\displaystyle g_{ij}(r,x) 
   = \stackrel{\scriptscriptstyle(0)}{g}_{ij}(x) + \frac{\ell^2}{r^2}\stackrel{\scriptscriptstyle(2)}{g}_{ij}(x) 
   + \frac{\ell^4}{r^4}\stackrel{\scriptscriptstyle(4)}{g}_{ij}(x)$} ,
\label{ae5}
\end{equation}
where $i=0,1$.  (Higher dimensions require an infinite series for ${g}_{ij}$, but the three 
dimensions the series terminates.)  The Einstein field equations reduce to \cite{Skenderis}
\begin{equation}
\stackrel{\scriptscriptstyle(2)}{g}\!{}^i{}_i = -\frac{\ell^2}{2}\stackrel{\scriptscriptstyle(0)}{R} ,\quad
\stackrel{\scriptscriptstyle(0)}{\nabla}_i \stackrel{\scriptscriptstyle(2)}{g}_{jk} 
    - \stackrel{\scriptscriptstyle(0)}{\nabla}_j \stackrel{\scriptscriptstyle(2)}{g}_{ik} = 0 ,\quad
\stackrel{\scriptscriptstyle(4)}{g}_{ij}=\frac{1}{4}\stackrel{\scriptscriptstyle(2)}{g}_{ik}%
     \stackrel{\scriptscriptstyle(0)}{g}\!{}^{kl}\stackrel{\scriptscriptstyle(4)}{g}_{lj} ,
\label{ae6}
\end{equation}
where the boundary metric $\stackrel{\scriptscriptstyle(0)}{g}_{ij}$ is used to raise and lower indices.
For a constant boundary metric, the general solution near conformal infinity is known \cite{Banadosa},
\begin{equation}
ds^2 = \frac{\ell^2}{r^2}dr^2 - 4G\ell(L^+du^2 + L^-dv^2) 
   - \left(\ell^2r^2 + \frac{16G^2L^+L^-}{r^2}\right)dudv ,
\label{ae7}
\end{equation}
where $u$ and $v$ are light cone coordinates, $L^+$ is an arbitrary function of $u$, and $L^-$ is 
an arbitrary function of $v$.

Note that solutions (\ref{ae7}) with different values of $L^\pm$ are physically inequivalent \cite{Banadosa}.
While they are related by diffeomorphisms \cite{Rooman}, these diffeomorphisms fall off too slowly
at infinity to be generated by the constraints (\ref{ad3}).  Benguria et al.\ \cite{Benguria} call these 
``improper'' gauge transformations; in (incomplete) analogy with \S\ref{large}, they are sometimes 
referred to as large diffeomorphisms.  In the present context, they are precisely the transformations $\xi$ 
and ${\bar \xi}$ in (\ref{ae3}) that generate the asymptotic symmetries of the spacetime.  Thus in 
contrast to the spatially closed case, spatially open (2+1)-dimensional vacuum solutions involve 
infinitely many boundary degrees of freedom.

Although they are not as well developed, similar results exist for the $\Lambda=0$ case \cite{Barnich,Gru}, 
which can be obtained from $\Lambda<0$ by a limiting process \cite{Bagchi}.  The asymptotic 
symmetry is then BMS${}_3$, the lower dimensional version of the Bondi-Metzner-Sachs symmetry of 
asymptotically flat (3+1)-dimensional spacetime at null infinity.   The asymptotic algebra becomes
\begin{align}
\{L[\xi],L[\eta]\} &= L[\eta\xi' - \xi\eta'] \nonumber\\
\{L[\xi],M[\eta]\} &= M[\eta\xi' - \xi\eta'] + \frac{c_{\hbox{\tiny\it LM}}}{4\pi}\int dz\left( \eta'\xi'' - \xi'\eta''\right)
\label{ae8}\\
\{M[\xi],M[\eta]\} &= 0 ,\nonumber
\end{align}
again with a central charge, $c_{\hbox{\tiny\it LM}} = \frac{3}{G}$.
 
\subsection{Two examples \label{examples}}

Let us consider two examples, one spatially closed and one open, that can serve as exemplars
when we consider quantization.  These are simple cases---the torus universe and the BTZ black 
hole---but they illustrate many of the crucial features.

\begin{enumerate}
\item{\bf The torus universe}\\[1ex]
Consider a vacuum spacetime, for simplicity with $\Lambda=0$, with topology $\mathbb{R}\times T^2$.
The torus $T^2$ has a fundamental group $\mathbb{Z}\oplus\mathbb{Z}$, so  the group of holonomies 
of the geometric structure is generated by two commuting elements of $\hbox{ISO}(2,1)$, determined 
up to conjugation.  If we choose standard Cartesian coordinates $(t,x,y)$ for Minkowski space, 
any such pair in the relevant component---see \S4.5 of \cite{Carlip_book} for details---can be conjugated 
to
\begin{align}
&\Lambda_1: (t,x,y)\rightarrow (t\cosh\lambda + x\sinh\lambda, x\cosh\lambda+t\sinh\lambda,y+a)
\nonumber\\
&\Lambda_2: (t,x,y)\rightarrow (t\cosh\mu + x\sinh\mu, x\cosh\mu+t\sinh\mu,y+b) ,
\label{af1}
\end{align}
with four real parameters $(\mu,\lambda,a,b)$.  A flat connection with these holonomies, unique
up to gauge transformations, is \cite{Carlip_tor}
\begin{alignat}{4}
&e^0 = e^{-t}dt \qquad && e^1 = adx+bdy \qquad &&e^2=e^{-t}(\lambda dx + \mu dy) \nonumber\\
&\omega^0 = 0 && \omega^1 = \lambda dx + \mu dy  && \omega^2= 0 ,
\label{af2}
\end{alignat}
so the symplectic form (\ref{ac6}) becomes
\begin{equation}
\Omega[(\delta e,\delta\omega)_{(1)},(\delta e,\delta\omega)_{(2)}]
   = \frac{1}{\kappa^2}\int \left(-\delta a^{(1)}\delta\mu^{(2)} + \delta b^{(1)}\delta\lambda^{(2)}
   - (1\leftrightarrow2)\right)dxdy ,
\label{af3}
\end{equation}
that is,
\begin{equation}
\{a,\mu\} = -\{b,\lambda\} = \kappa^2  .
\label{af4}
\end{equation}
The geometry (\ref{af2}) can also be obtained explicitly as a quotient space of a 
region of (2+1)-dimensional Minkowski space by the action of the group generated by 
$\langle\Lambda_1,\Lambda_2\rangle$ \cite{Carlip_tor}.

It is straightforward to translate the metric obtained from (\ref{af2}) into the ADM formalism.  
The York time is $T=e^t$, so the surfaces of constant $t$ are already constant mean
curvature slices.  Restricting the metric to such a slice, one can read off the moduli $m_\alpha$,
which are now just the standard modulus $\tau$ and its complex conjugate.  The
conjugate momenta (\ref{ad6}) are also easily calculable \cite{Carlip_tor}:
\begin{equation}
\tau = \tau_1 + i\tau_2 = \left(a + \frac{i\lambda}{T}\right)^{-1}\left(b + \frac{i\mu}{T}\right), \quad
p = p_1 + ip_2 = -\frac{iT}{2\kappa^2}\left(a - \frac{i\lambda}{T}\right)^2 ,
\label{af5}
\end{equation}
where, from (\ref{af4}), $\{\tau,{\bar p}\}=2$, as expected.  For the torus, Eqn.\ (\ref{ad6}) for 
the ADM scale factor can be solved explicitly, yielding a Hamiltonian
\begin{equation} 
H = \frac{1}{2\kappa^2}\frac{\tau_2}{T}\left(p_1{}^2 + p_2{}^2\right)^{1/2} 
   = \frac{1}{2\kappa^2}\frac{a\mu - \lambda b}{T} ,
\label{af6}
\end{equation}
and the brackets (\ref{af4}) give the correct Hamilton's equations of motion for $\tau$ and $p$.

\item{\bf The BTZ black hole}\\[1ex]
Next consider  a spacetime $\mathbb{R}\times\mathbb{R}^2$ with $\Lambda=-1/\ell^2<0$ and 
anti-de Sitter boundary conditions (\ref{ae2}).  In 1992, Ba{\~n}ados,
Teitelboim, and Zanelli showed that the vacuum field equations  admit a black hole solution,
 a special case of (\ref{ae7}) with
\begin{equation}
L^\pm = \frac{(r_+\pm r_-)^2}{16G\ell} .
\label{ag1}
\end{equation}
The corresponding Chern-Simons connection (\ref{ac3b}) is
\begin{equation}
A^{(+)} = \left(\begin{array}{cc} \frac{1}{2r}dr & -\frac{4GL^+}{\ell r} du\\[1ex]
                                                 -rdu &  -\frac{1}{2r}dr \end{array}\right) ,\qquad
A^{(-)} = \left(\begin{array}{cc} -\frac{1}{2r}dr & -rdv\\[1ex]
                                                 -\frac{4GL^-}{\ell r} &  \frac{1}{2r}dr \end{array}\right) ,
\label{ag2}
\end{equation}                                                 
with $\hbox{SL}(2,\mathbb{R})\times\hbox{SL}(2,\mathbb{R})$ holonomies that can be
conjugated to \cite{Carlip_bh}
\begin{equation}
\rho_L = \left(\begin{array}{cc} e^{\pi(r_+-r_-)/\ell} & 0\\ 0 & e^{-\pi(r_+-r_-)/\ell}\end{array}\right), \quad
\rho_R = \left(\begin{array}{cc} e^{\pi(r_++r_-)/\ell} & 0\\ 0 & e^{-\pi(r_++r_-)/\ell}\end{array}\right) .
\label{ag3}
\end{equation}
As in the compact case, these holonomies describe a geometric structure, and the BTZ spacetime
can be obtained as a quotient of the universal cover of anti-de Sitter space by the group generated
by $\langle\rho_L,\rho_R\rangle$ \cite{BHTZ}.

The existence of a black hole spacetime is rather surprising, since all solutions have constant negative 
curvature, but the BTZ solution is clearly a black hole \cite{BTZ,Carlip_bh}: it has an event horizon at
$r=r_+$ (which is also a Killing horizon), an inner Cauchy horizon at $r=r_-$, a Penrose diagram 
almost identical to that of the Kerr-AdS black hole, and an ADM mass and angular momentum
\begin{equation}
M = \frac{r_+{}^2 + r_-{}^2}{8G\ell^2}, \quad J = \frac{r_+r_-}{4G\ell}  .
\label{ag4}
\end{equation}

It would also be interesting to look at the BTZ solution in the ADM formalism of \S\ref{ADM}.  
Unfortunately, though, the presence of a horizon makes York time slicing $T=-K$ much more
complicated, and very little has been achieved in this direction.
\end{enumerate}

\section{Quantum gravity: spatially compact case}

We can now turn to the problem of quantizing the classical theory.  With no propagating
excitations or gravitons, quantum gravity in 2+1 dimensions is, of course, not a very realistic
model.  But it is a valuable test bed for programs of quantization, such as loop quantum gravity
and covariant canonical quantization, and it might also provide insight into some of the deep 
conceptual problems of quantum gravity, such as the notorious ``problem of time'' \cite{Kuchar}.  

Here we provide a brief introduction to some of the principal work, starting with the spatially 
compact case.  This will necessarily be a rather cursory overview; in particular,
the important but enormous topic of discrete state sum models will receive only a brief
summary.

There is, of course, no universal algorithm for ``quantizing'' a classical theory.  In general,
we want to map functions on phase space to operators acting on a Hilbert space, with 
commutators
\begin{equation}
\{\,f,g\}  \rightarrow \frac{1}{i\hbar} [\,{\hat f},{\hat g}] .
\label{b1}
\end{equation}
In general, though, no consistent mapping of this form exists for the entire phase space;
one must either apply the mapping to a ``big enough'' subset of phase space functions or view
(\ref{b1}) as the first term in an expansion \cite{Ishamb,deform}.  Different choices may lead to
different quantum theories.  Similarly, different choices of the Hilbert space and inner product may
lead to different theories, and it is worthwhile to explore a range of possibilities.

\subsection{Reduced phase space quantization \label{reduced}}

For a spatially closed universe, (2+1)-dimensional gravity has a finite number of physical
degrees of freedom, and the most straightforward approach to quantization is to isolate
these and find a corresponding quantum theory.  The ADM formalism of \S\ref{ADM}
provides a setting to do this.  The classical action reduces to (\ref{ad7}), that is, to an ordinary 
looking finite dimensional quantum mechanical action with a Hamiltonian determined by 
(\ref{ad5}).  Quantization is ``obvious'': we impose standard Heisenberg commutators,
represent momenta as derivatives, and write down the usual Schr{\"o}dinger equation,
\begin{equation}
[{\hat m}_\alpha,{\hat p}^\beta] = i\hbar\delta^\alpha_\beta, \qquad 
i\hbar\frac{\partial\psi(m_\alpha,T)}{\partial T} 
   = {\hat H}\left({\hat m}_\alpha,{\hat p}^\beta,T\right)\psi(m_\alpha,T)  .
\label{ba1}
\end{equation}

Of course, life is not so simple.  For a space of genus $g>1$, Eqn.\ (\ref{ad5}) for the scale 
factor cannot be solved in closed form, and any perturbative solution will be plagued with terrible
operator ordering ambiguities.  For the torus universe, though, the Hamiltonian is given exactly 
by (\ref{af6}).  Upon quantization we obtain, up to ordering ambiguities,
\begin{equation}
{\hat H} = \frac{\hbar^2}{2\kappa^2}T^{-1}\Delta_0^{1/2}, \qquad 
\Delta_0 = -\tau_2{}^2\left(\frac{\partial^2\ }{\partial\tau_1{}^2} + \frac{\partial^2\ }{\partial\tau_1{}^2}\right) ,
\label{ba2}
\end{equation}
where we take the positive square root of $\Delta_0$ to ensure that $\hat H$ is bounded below.
The operator $\Delta_0$ is the Maass Laplacian for weight zero forms; it is invariant under the
torus mapping class group, and has been extensively studied by mathematicians \cite{Iwaniec}.
Similar results can be found for $\Lambda\ne0$ \cite{CarlipNelson}.  

The passage from (\ref{af6}) to (\ref{ba2}) required a choice of operator ordering,  While this
choice is not unique, it is highly constrained by the demand that the mapping class group
$\mathcal{D}(\Sigma)$ act unitarily as a symmetry \cite{Carlip_modular}.  For this, the Laplacian $\Delta$
must act on automorphic forms of weight $n/2$, where $n$ is an integer, and the corresponding
Maass Laplacians are then fixed.  These, again, have been widely studied by mathematicians
\cite{Fay,Maass}.  (If one starts with the Chern-Simons formalism, it seems that the most natural 
choice is $n=1/2$ \cite{Carlip_Dirac}.)  

\subsection{The Wheeler-DeWitt equation \label{WdW}}

Even when it can be carried out in full, the reduced phase space quantization of \S\ref{reduced} 
has some peculiarities.  It requires a classical choice of time, and it is not obvious that a
different choice would give an equivalent quantum theory.  Indeed, it is not even clear how
one would ask questions about observables on a time slice that was not one of constant
York time.

The traditional alternative is Dirac quantization \cite{DeWitt}, in which the constraints (\ref{ad3}) are 
not solved classically, but are rather imposed as operator equations:
 \begin{align}
&\pi^{ij} \rightarrow \frac{\hbar}{i}\frac{\delta\ }{\delta q_{ij}} \nonumber\\[1ex]
&\left\{\frac{\hbar}{i}{}^{\scriptscriptstyle(2)}\nabla_j\frac{\delta\ }{\delta q_{ij}}\right\}\Psi[q] = 0
\label{bb1}\\[1ex]
&\left\{\frac{2\hbar^2\kappa^2}{\sqrt{q}} q_{ij}q_{kl}\left(\frac{\delta\ }{\delta q_{ik}}\frac{\delta\ }{\delta q_{jl}}
    -\frac{\delta\ }{\delta q_{ij}}\frac{\delta\ }{\delta q_{ij}}\right)
   + \frac{1}{2\kappa^2}\sqrt{q}({}^{\scriptscriptstyle(2)}\!R - 2\Lambda)\right\}\Psi[q] = 0 .
\label{bb2}
\end{align}
Eqn.\ (\ref{bb1}), the operator form of the momentum constraints, requires that $\Psi[q]$ be
invariant under spatial diffeomorphisms, while (\ref{bb2}), the operator Hamiltonian constraint,
is the (2+1)-dimensional Wheeler-DeWitt equation.  It is almost certainly necessary to 
further gauge fix the inner product on the space of solutions \cite{Woodard}, and the
product of functional derivatives must be regulated \cite{Woodardb}, but then, in principle, 
solutions should give us quantum amplitudes for spatial geometries without a prior choice of time
slicing.

The main thing we have learned from 2+1 dimensions is how hard this is \cite{Carlip_WDW}.  
Because their geometrical meaning is clear, the momentum constraints (\ref{bb1}) are usually
dismissed as trivial.  But their solutions are necessarily nonlocal, and when inserted back
into the Wheeler-DeWitt equation (\ref{bb2}), they lead to unmanageable nonlocal terms.  One
can try to expand in known  invariant quantities \cite{Banks,Leut}---spatial volume, integrated 
curvature, etc.---but this is not a well controlled expansion, and it quickly becomes very  complicated.

There is also an intermediate approach between reduced phase space quantization and the 
Wheeler-DeWitt equation.  One can (somewhat arbitrarily) \emph{partially} fix the gauge and   
apply Dirac quantization to the remaining simplified action.  For the torus, for instance, imposing 
the York time gauge and solving the momentum constraints while retaining the Hamiltonian 
constraint leads to a reduced Wheeler-DeWitt equation that is essentially the square of the 
Schr{\"o}dinger equation of \S\ref{reduced} \cite{Hosoya}.

\subsection{Chern-Simons quantization \label{CSq}}

In Chern-Simons quantization, the fundamental variables are the holonomies of \S\ref{CS}, with
the symplectic structure (\ref{ac6}).   Nelson, Regge and Zertuche have constructed a small but 
complete set of holonomies on a surface of arbitrary genus that form a closed algebra 
\cite{NelReg,NelRegZer,Nelsonc,Nelsond}, and analyzed the commutators obtained by the
procedure (\ref{b1}), finding the first indications of the appearance of quantum groups.  This
work has been extended by a number of authors \cite{Buff,Meusz,Balles,Noui,Dupuis}, employing 
the combinatorial quantization methods introduced Fock and Rosly \cite{Fock} and Alekseev 
\cite{Alekseev1,Alekseev2,Alekseev3}.  While the quantum theory is not yet fully understood, it 
seems clear that quantum groups, and in particular Drinfel'd doubles, will play a critical role.
(For a nice review, see \cite{Schroers}.)  

As in reduced phase space quantization, the torus universe of \S\ref{examples} provides a particularly
simple example.   Let us specialize to the case $\Lambda = -1/\ell^2<0$, so $G_\Lambda
\approx\hbox{SL}(2,\mathbb{R})\times\hbox{SL}(2,\mathbb{R})$.  Let $\gamma_1$ and $\gamma_2$ 
be two simple closed curves that generate $\pi_1(T^2)$, with $\gamma_{12} = \gamma_1\circ\gamma_2$,
and denote the corresponding holonomies as $R^\pm_1$, $R^\pm_2$, and $R^\pm_{12}$.
Then the quantization map (\ref{b1}) gives the commutators
\begin{equation}
{\hat R}_1^{\pm}{\hat R}_2^{\pm}e^{\pm i \theta}   - {\hat R}_2^{\pm}{\hat R}_1^{\pm} e^{\mp i \theta}=
   \pm 2i\sin\theta\, {\hat R}_{12}^{\pm} \quad \hbox{\it and cyclical permutations}
\label{bc1}
\end{equation}
with the added relation
\begin{equation}
{\hat F}^{\pm}  = 1-{\tan}^2\theta - e^{\pm 2i\theta} \left( ({\hat R}_1^\pm)^2 + ({\hat R}_{12}^\pm)^2\right) 
  - e^{\mp 2i\theta} ({\hat R}_2^\pm)^2 
  + 2e^{\pm i\theta}\cos\theta {\hat R}_1^\pm \hat R_2^\pm {\hat R}_{12}^\pm = 0 ,
\label{bc2}
\end{equation}
where $\tan\theta= -{\hbar\kappa^2/8\ell}$.  The relations ${\hat F}^\pm=0$ are the quantum version
of the Mandelstam identities \cite{Mandelstam}, relations among holonomies that follow from 
$\hbox{SL}(2,\mathbb{R})$ identities.  The algebra (\ref{bc1})  is not a Lie algebra, but it is closely related 
to the algebra of the quantum group $U_q(\mathit{sl}(2))$ with $q=\exp{4i\theta}$ and a $q$-Casimir 
$F$ \cite{NelRegZer}.  Similar results exist for $\Lambda\ge0$; for details, see\cite{CarlipNelson}.

\subsection{Loop quantum gravity}

Like Chern-Simons quantization, loop quantum gravity starts with a space of holonomies, but
it chooses a very different Hilbert space.  Let us again assume a spacetime topology 
$\mathbb{R}\times\Sigma$, and for simplicity take $\Lambda=0$.  Let $\mathscr{ J}_a$ be
the generators of $\hbox{SL}(2,\mathbb{R})$, and define
\begin{equation}
\rho_0[\gamma,x] = P\exp\left\{\int_\gamma \omega^a\mathscr{ J}_a \right\} ,
\label{bd1}
\end{equation}
the holonomy matrix of the spin connection $\omega$ around the curve $\gamma$ with base
point $x$.  The Ashtekar-Smolin-Rovelli loop variables \cite{Ash_loop} are then
\begin{equation}
\mathcal{T}^0[\gamma] = \frac{1}{2}\mathrm{Tr}\rho_0[\gamma,x], \quad
\mathcal{T}^1[\gamma] =  \int_\gamma 
\mathrm{Tr}\left\{ \rho_0[\gamma, x(s)]\, e^a(\gamma(s)){\mathcal J}_a \right\} .
\label{bd2}
\end{equation} 
Classically, these obey a closed Poisson algebra similar to that of the Chern-Simons holonomies 
of \S\ref{CSq}, again with Mandelstam constraints \cite{Mandelstam} coming from relations
among elements of $\hbox{SL}(2,\mathbb{R})$; see, for instance, \S4.7 of \cite{Carlip_book}.
The operators corresponding to $\mathcal{T}^0$ and $\mathcal{T}^1$ become the fundamental 
quantum variables of loop quantum gravity.

The  Mandelstam constraints again complicate the quantum theory, and in 3+1 
dimensions, holonomies around loops are usually replaced by (unconstrained) spin networks.
For the torus universe, though, the original loop quantization can be carried out in full.  
Any homotopy class of curves on the torus is completely characterized by a pair $(m,n)$ 
of winding numbers, so we can write the fundamental operators as $\mathcal{\hat T}^0[m,n]$
and $\mathcal{\hat T}^1[m,n]$.  We define a vacuum state $|0\rangle$ annihilated by the 
$\mathcal{\hat T}^1[m,n]$, and treat the $\mathcal{\hat T}^0[m,n]$ as raising operators,
 \begin{align}
\mathcal{\hat T}^0[m,n] |p,q\rangle &= \phantom{-} \frac{1}{2} \left(
   |m+p,n+q\rangle + |m-p,n-q\rangle\right) \nonumber\\
\mathcal{\hat T}^1[m,n] |p,q\rangle &= -\frac{i}{8} (mq-np) \left(
   |m+p,n+q\rangle - |m-p,n-q\rangle\right) .
\label{bd3}
\end{align}
It may be checked that (\ref{bd3}) gives a representation of the loop algebra, and is thus a
legitimate quantization of the theory.  It is \emph{not}, however isomorphic to Chern-Simons
quantization, at least without some rather contrived manipulations \cite{Marolf}.  In fact, the
natural mapping between the two sends a dense subspace of Chern-Simons states to zero.  
(For more details, see \S7.3 of \cite{Carlip_book}.)

A more formal approach to loop quantization has also been explored, although mainly for
`` Euclidean gravity,'' that is, gravity with a Riemannian signature spacetime metric \cite{NouiPerez}.  
The fundamental operators are again the holonomies and fluxes (\ref{bd2}), now with a compact 
gauge group $\hbox{SU}(2)\times\hbox{SU}(2)$.  These act on an auxiliary Hilbert space built 
out of functions of holonomies on graphs, essentially spin networks, and the constraints can be 
imposed as projectors in the inner product.  The result has an interpretation as a sum over spin 
foams, and is related to the Ponzano-Regge model of \S\ref{Lat}.  A related analysis of the 
quantized constraint algebra leads to the appearance of a quantum group structure similar to 
that seen in Chern-Simons quantization \cite{Cianfrani}.

\subsection{Covariant canonical quantization}

As noted in \S\ref{CS}, the Chern-Simons holonomies are essentially the same as the holonomies
of the geometric structure.  In particular, in the relevant connected component of the moduli space
$\mathcal{M}$, they are coordinates on the space of vacuum spacetimes, unique up to a possible
action of the mapping class group.  The Chern-Simons quantization of \S\ref{CSq} is thus an example 
of ``quantizing the space of solutions,'' or covariant canonical quantization \cite{Bombelli,Crn,Crnb,Lee}. 

Phase space, the usual setting for canonical quantization, is ordinarily defined as the space of
initial data on some Cauchy surface $\Sigma_t$.  Suppose, though, that we are dealing with a 
system with a well posed initial value problem.  Then each point in phase space  determines a
classical solution, and, conversely, the restriction of any solution to $\Sigma_t$ determines a point 
in phase space.  This provides a homeomorphism between phase space and the space of solutions,  
and requiring it to be a diffeomorphism determines a topology on the space of solutions (up to some  
subtleties \cite{Woodhouse}).  This map is, moreover, a symplectomorphism, taking the standard
symplectic structure on phase space to the covariant symplectic structure of \cite{Lee}.

This equivalence suggests a strategy to address the ``problem of time'' in quantum gravity \cite{CarHu}.
Covariant canonical quantum theory has no time dependence---one is, after all, quantizing entire 
histories---and it is not obvious how to construct interesting time-dependent observables.  For
(2+1)-dimensional gravity we know the classical answer, though: use the holonomies to 
construct a spacetime, choose a favorite time slicing, and read off the metric on a slice $\Sigma_t$.   
At least for  $M = \mathbb{R}\times T^2$, this procedure translates directly to the quantum theory.  
As we saw in \S\ref{examples}, the ADM moduli and momenta on a slice of constant York time can be
written explicitly in terms of the holonomies, with the time $T$ appearing as a parameter.  As 
operator equations, these become examples of what Rovelli calls ``evolving constants of
motion,'' one-parameter families of diffeomorphism invariant operators with a parameter 
determined by a classical choice of time slicing \cite{Rovelli}.

Quantizing the Poisson brackets (\ref{af4})---essentially Chern-Simons quantization---gives us 
time-independent wave functions $\Psi(\lambda,\mu)$.  But the moduli $(\tau(T),{\bar\tau}(T))$ of 
(\ref{af5}) now become ``time''-dependent operators.  We can diagonalize these operators to 
obtain a transition matrix $\left\langle\mu,\lambda|\tau,{\bar\tau};T\right\rangle$, and use this to define 
a $\tau$ representation of the wave functions \cite{Carlip_tor,Carlip_modular,Carlip_Dirac,CarlipNelson},
\begin{equation}
{\tilde\Psi}(\tau,{\bar\tau},T) 
   = \int\!d\lambda d\mu\, \left\langle\mu,\lambda|\tau,{\bar\tau};T\right\rangle \Psi(\lambda,\mu) .
\label{be1}
\end{equation}
Up to the ordering ambiguities discussed in \S\ref{examples}, these are precisely
the reduced phase space wave functions of \S\ref{reduced}, and the choice of a representation
of the mapping class group completely determines the operator ordering, the inner product,
and the normalization of $\left\langle\mu,\lambda|\tau,{\bar\tau};T\right\rangle$.

If we had chosen a different classical time slicing, we would have obtained a different
set of moduli ${\tilde\tau}({\tilde T})$, a different transition matrix, and different ``time''-dependent
wave functions.  But the descriptions would be at least formally consistent, since all would come 
from the same underlying covariant quantized wave functions.  Extending these results to higher 
genus spacetimes is technically difficult, but  perhaps not impossible.

\subsection{Path integrals}

A standard alternative to canonical quantization is the Feynman path integral.   The typical setting 
for a path integral in quantum gravity is a manifold $M$ with boundary $\partial M$ and a set of 
geometric variables $\mathfrak{g}$ with boundary values $\mathfrak{g}|_{\partial M} = \mathfrak{h}$.
The path integral is then
\begin{equation}
Z[\mathfrak{h}] = \int\limits_{\mathfrak{g}|_{\partial M} = \mathfrak{h}}\![d\mathfrak g\,]
   \exp\left\{\frac{i}{\hbar}I[\mathfrak{g}]\right\} .
\label{bf1}
\end{equation}
For a boundary $\partial M$ with two or more components, this expression gives a transition amplitude 
between initial and final boundary values; for one component, it gives an amplitude for nucleation of a 
universe from ``nothing''; and for a manifold without boundary, it gives a partition function (thermal 
if one fixes a periodicity in imaginary time).  In the metric formalism, the variables $\mathfrak{g}$ can 
be the spacetime metric $g$ or the set $(q,\pi,N,N^i)$ in the ADM decomposition; in the first order 
formalism they can be $(e,\omega)$ or the Chern-Simons connection $A$.
 
\begin{enumerate}
\item{\bf The metric path integral}\\[1ex]
The simplest path integral approach starts with the pair $(q,\pi)$ and the action (\ref{ad2}).  The lapse
and shift functions $N$ and $N^i$ occur as Lagrange multipliers, and give delta functions of the 
constraints $\mathcal{H}$ and $\mathcal{H}_i$.  One might therefore expect the path integral to be
equivalent to the canonical quantization of \S\ref{ADM}, in which these constraints are solved before
quantization.  This is, indeed, the case \cite{Carlip_canpi,Seriu}.

If, instead, the variables are taken to be the spacetime metric with the action (\ref{aa1}), the path
integral becomes technically much more difficult.  For $\Lambda<0$, however, the one-loop
contribution for asymptotically anti-de Sitter space has been computed, after analytic continuation
to Riemannian signature \cite{MaloneyXi}.  It has been argued that no higher order corrections appear \cite{MaloneyWitten}, and this has been directly checked at two and three loop order, where some
rather unobvious cancellations occur \cite{Leston}.  The one-loop calculation has been repeated 
for asymptotically flat gravity with $\Lambda=0$ \cite{BarnichMaloney}, and for lens spaces with
$\Lambda>0$ \cite{Maloney_lens}.
  
\item{\bf The first order formalism}\\[1ex]
The first order formalism and its Chern-Simons cousin have been especially useful settings for the 
path integral.   In a pair of seminal papers \cite{Wittena,Wittenb}, Witten showed that the path
integral with $\Lambda=0$ could be expressed in terms of a set of determinants, in a combination  
that had an elegant topological interpretation as the Ray-Singer torsion.  The partition function for 
a closed manifold $M$ takes the form
\begin{equation}
Z_M  = \int_{\mathcal{M}} \!d{\bar\omega}\,d{\bar e}\,
\frac{|\mathop{det}\Delta_{\bar\omega}^{(3)}|_{\phantom{\bar\omega}}^{3/2}
  |\mathop{det}\Delta_{\bar\omega}^{(1)}|_{\phantom{\bar\omega}}^{1/2}}
  { |\mathop{det}\Delta_{\bar\omega}^{(2)}|}
\label{bf2}
\end{equation}
where $({\bar\omega},{\bar e})$ are a flat $\hbox{ISO}(2,1)$ connection---that is, an element of the
moduli space (\ref{ab2})---and 
$\Delta_{\bar\omega}^{(n)} = D_{\bar\omega}*D_{\bar\omega}* +  *D_{\bar\omega}*D_{\bar\omega}$ 
is the gauge-covariant Laplacian acting on $n$-forms.  The extension to manifolds with boundary is
relatively straightforward \cite{Wu,CarlipCosgrove}.  For a spatially compact product 
manifold $\mathbb{R}\times\Sigma$, the transition amplitude is nonzero only if the holonomies
of $\omega$ on the initial and final surfaces agree, as one would expect from canonical quantization.

\item{\bf Topology change}\\[1ex]
Canonical quantization normally requires a spacetime $\mathbb{R}\times\Sigma$ with a fixed spatial 
topology.  Path integral approaches, on the other hand, allow topology changes and sums over
topologies \cite{Wittenb}.  The possibilities are restricted; if  the initial and final surfaces are required to 
be spacelike and nondegenerate, they must have equal Euler numbers \cite{Amano}.  But within this 
constraint, examples of topology changing amplitudes have been explicitly calculated \cite{CarlipCosgrove}, 
although with divergences arising from  the integration over moduli in (\ref{bf2}) that are not 
completely understood.

For $\Lambda\ne0$, the path integral in the first order formalism becomes harder.  The
Chern-Simons path integral is well understood for compact gauge groups \cite{Wittenc}, but much
less so in the noncompact case \cite{Wittend}, despite some interesting recent progress \cite{Wittene}.  
Moreover, many of the most interesting configurations---for instance, the nucleation of a 
universe from nothing---admit no Lorentzian metrics.  Consequently, much of the work on topology 
change and sums over topology has been carried out in the context of Euclidean quantum gravity,
in which the Lorentzian metrics in the path integral are replaced by metrics with Riemannian signature.  

For $\Lambda>0$, this switch in signature amounts to changing the Chern-Simons gauge group 
to $\hbox{SO}(4)$.  The three-manifolds that admit constant positive curvature Riemannian metrics
are completely classified, and a partial sum over topologies---specifically over lens spaces---was
carried out in \cite{Carlip_sum}, corrected in \cite{Maloney_lens}.  The resulting sum diverges, in 
a way that does not have an obvious cure.  For $\Lambda<0$, the three-manifolds of constant
curvature are not completely classified, but again it can be shown that the sum over topologies
diverges \cite{Carlip_sum}.  

It is also interesting to consider the Hartle-Hawking wave function, the function 
$\psi[\mathfrak{h}]$ coming from the path integral (\ref{bf1}) for a manifold with a single boundary.
Such aa wave function may be interpreted as an initial wave function for a universe with no past
boundary, that is, a universe born from ``nothing'' \cite{HH}.  For $\Lambda<0$, the sum over 
topologies has bee considered in \cite{Carlip_HH}.  The sum again diverges, but probably only 
at certain specific values of the boundary metric $h$ corresponding to rigid boundary surfaces.  
If this behavior is confirmed, it suggests that a properly normalized wave function 
may be infinitely peaked at a small number of spatial metrics, giving the path integral unexpected
predictive value.
\end{enumerate}

\subsection{Discrete approaches \label{Lat}}

The idea of a discrete quantization of (2+1)-dimensional gravity dates back to the early 1960s,
with Regge's exploration of the Einstein-Hilbert action for a simplicial complex \cite{Regge0}.
For a simplicial three-manifold with edges of length $\ell_e$, the Riemannian signature action is
\begin{equation}
I_{\mathit{Regge}}
   = 2\sum_{\hbox{\scriptsize edges\,$e$}}\delta_e \ell_e ,
\label{bg1}
\end{equation}
where $\delta_e$ is the conical deficit angle at the edge labeled by the index $e$.  As is
customary, we choose units $2\kappa^2=1$; in the quantum theory this amounts to saying
that lengths are measured in units of $16\pi\ell_p$, where $\ell_p= \hbar G$ is the
(2+1)-dimensional Planck length.  A similar but slightly more complicated expression exists 
for Lorentzian signature, where some deficit angles  become Lorentzian boosts \cite{Barrett}.  
An analogous action exists in any dimension, but in higher dimensions, simplicial manifolds are 
typically approximations, while in 2+1 dimensions they can be exact. 

As Ponzano and Regge observed \cite{Ponzano}, the action (\ref{bg1}) can be expressed in
terms of Wigner-Racah $6j$ symbols. For a single tetrahedron with edge lengths 
$\ell_i = \frac{1}{2}(j_i + \frac{1}{2})$ and volume $V$, 
\begin{equation}
\exp\{{\pi i}\sum_{i=1}^6 j_i\} \left\{ 
  \begin{array}{ccc} 
  j_1 & j_2 & j_3\\ j_4 & j_5 & j_6 \end{array} \right\}
  \sim
\frac{1}{\sqrt{6\pi V}}\left[ 
  \exp\left\{i\left(I_{\mathit{Regge}}+\frac{\pi}{4}\right)
  \right\} + 
  \exp\left\{-i\left(I_{\mathit{Regge}}+\frac{\pi}{4}\right) 
  \right\}  \right] 
\label{bg2}
\end{equation}
in the limit of large $j$ \cite{Roberts}.  For a manifold made up of many simplices, the exponential of 
the Regge action thus looks like a product over such $6j$ symbols, and the gravitational path integral 
is a sum of such products  \cite{Hasslacher,Ooguri}.  Such a construction is an example of a state 
sum model \cite{Barrettb}, and forms part of an very large and active field , which largely 
falls outside the scope of this article.

The sum over edge lengths the Ponzano-Regge model diverges---perhaps because of a
residual diffeomorphism invariance in the Regge action \cite{FreidelL}---and must be regulated.
The Turaev-Viro model \cite{Turaev} does so by replacing the spins $j$ in (\ref{bg2}), viewed as
representations of $\hbox{SU}(2)$, by representations of the quantum group $U_q(\mathit{sl}(2))$,
with $q = \exp\left\{\frac{2\pi i}{k+2}\right\},\ k\in\mathbb{Z}$.  The number of such representations 
is finite, and the Ponzano-Regge amplitude no longer requires regularization.  One effect of this
change is to introduce a cosmological constant \cite{Mizoguchi},
\begin{equation}
\Lambda = \left(\frac{4\pi}{k}\right)^2 ,
\label{bg3}
\end{equation}
which vanishes as $k\rightarrow\infty$.   The construction of 
physical states as functions of boundary edge lengths is summarized in \S11.2 of \cite{Carlip_book}.
The Turaev-Viro model and its later descendants have been connected to loop quantum gravity 
and spin foams \cite{NouiPerez,Hasslacher,Rovellix,Pran}, Chern-Simons 
quantization \cite{Oogurib,Freidelx} and a broad range of topics in mathematics, from three-manifold
invariants to knot theory to quantum groups and their representations \cite{Turaev_book,Barrettc}.

In a radically different strategy, the path integral (\ref{bf1}) may be approximated as
a sum over simplicial manifolds with the Regge action.  Two variations are common: one may fix
a triangulation and sum over the edge lengths $\ell_e$, or hold the lengths fixed and sum over 
inequivalent triangulations.  While the first alternative has received a bit of attention \cite{Hamber}, 
the second---commonly known as ``dynamical triangulations''---has been more popular, especially 
with the advent of ``causal dynamical triangulations'' \cite{Loll}.  A causal dynamical triangulation is 
a triangulation with a direction of time, imposed by choosing a (topological) time-slicing and 
restricting to simplices that fill in the space between adjacent slices.  The Regge action in 2+1 
dimensions depends only on the numbers of various types of simplices \cite{Loll}---see 
\cite{Cooperman} for the inclusion of boundary terms---and one can apply standard Monte Carlo 
methods to sum over configurations.  Two phases appear, a weakly coupled phase that looks very 
much like de Sitter space and a strongly coupled phase that is highly nonclassical \cite{Lollb}.  
Near the Planck scale, on the other hand, the behavior is always nonclassical: the spectral 
dimension of the spacetime falls from $d=3$ at large distances to $d\approx2$ at small distances 
\cite{Benedettic}.

\section{Quantum gravity: asymptotia}

As we saw in  \S\ref{open},  (2+1)-dimensional gravity with asymptotically anti-de Sitter
boundary conditions acquires an infinite set of new degrees of freedom at infinity.  These edge
modes have been described as ``would-be gauge degrees of freedom'' \cite{Carlip_wouldbe}:
they are related by diffeomorphisms, which would normally make them nonphysical, but the required
diffeomorphisms fall off too slowly at infinity to be true invariances, functioning rather as asymptotic 
symmetries.  Most of the work on (2+1)-dimensional quantum gravity with spatially open topologies
has focused on the effort to quantize these edge modes, in order to develop a theory of ``boundary
gravitons.''

The natural starting point is the asymptotic symmetry algebra (\ref{ae3}).  This is easily quantized: 
the Poisson brackets become commutators, and the substitutions
\begin{equation}
L[\xi] \rightarrow \frac{{\hat L}[\xi]}{\hbar}, \quad c\rightarrow\frac{c}{\hbar}
\label{c1}
\end{equation}
(and corresponding substitutions for $\bar L$) preserve the structure of the algebra.  This implies
that the boundary modes are described by a conformal field theory with central charges (\ref{ae4}).
 
Support for this conclusion comes from an important result of Cardy \cite{Cardy,Cardyb}, who showed 
that for a unitary conformal field theory, the asymptotic behavior of the density of states has an 
exceptionally simple form.  Denote eigenvalues of the zero modes $L_0$ and ${\bar L}_0$, the 
``conformal weights,'' by $\Delta$ and $\bar\Delta$, and let $\Delta_0$ and ${\bar\Delta}_0$ be their 
lowest  values, usually but not always zero.  Then for $\Delta$ and ${\bar\Delta}$ large, the density 
of states $\rho(\Delta,{\bar\Delta})$ has the asymptotic form
\begin{equation}
\ln\rho(\Delta,{\bar\Delta}) \sim 2\pi\sqrt{\frac{c_{\hbox{\tiny\it eff}}\Delta}{6}} 
    + 2\pi\sqrt{\frac{{\bar c}_{\hbox{\tiny\it eff}}{\bar \Delta}}{6}} ,
\label{c2}
\end{equation}
 where the ``effective central charges'' are
\begin{equation}
c_{\hbox{\tiny\it eff}} = c-24\Delta_0, \quad 
{\bar c}_{\hbox{\tiny\it eff}} = {\bar c}-24{\bar\Delta}_0 .
\label{c3}
\end{equation}
The derivation of (\ref{c2}) uses the modular invariance of the torus partition function; a proof can
be found in \cite{Carlipg}, with higher order corrections given in \cite{Carliph,Birm,Farey,Ubaldo}.
In 1998, Strominger \cite{Strominger} and Birmingham, Sachs, and Sen \cite{BSS} independently 
pointed out that if one takes this expression and inserts the Brown-Henneaux central charges
(\ref{ae4}) and the classical conformal weights (\ref{ag1}) for the BTZ black hole, one obtains
an entropy
\begin{equation}
S = \frac{2\pi r_+}{4\hbar G} ,
\label{c4}
\end{equation}
exactly the standard Bekenstein-Hawking entropy.

While this result strongly suggests that BTZ black hole entropy counts states of boundary fields,
it does not tell us what those states are.  The search for a quantum conformal field theory of  
edge modes is very active, but not yet completely successful \cite{Carlip_conf}.  In 
both the Chern-Simons \cite{Coussaert} and the metric \cite{Carlip_dyn} formalisms, the   
gravitational action with appropriate boundary terms induces a conformal field theory described
by the Liouville action---or perhaps a related model such as a ``Virasoro TQFT'' \cite{Collier} or
a generalization to incorporate projective structures \cite{Krasnov}---at the asymptotic
boundary.  The relation to ``would-be gauge degrees of freedom'' is especially clear in the metric
version: the boundary degrees of freedom are precisely the parameters labeling diffeomorphisms
that fail to fall off fast enough at infinity.  But while the resulting Liouville theory has the correct
central charge, the relevant states almost certainly fall in the ``nonnormalizable sector'' \cite{Seiberg},
whose quantization is very poorly understood. 

An alternative approach is to compute a thermal partition function by summing over classical saddle
points (with periodic imaginary time) and their first order quantum corrections, to see what that tells us 
about the boundary theory \cite{Witten_re,MaloneyWitten,Keller}.  With periodic time, the conformal 
boundary is a torus, so the partition function should be an automorphic function, again allowing the
use of powerful mathematical tools.  The result, however, is not the partition
function for any physically sensible theory: it cannot be written as a positive sum of exponentials, 
$\mathrm{Tr} e^{-\beta H}$, and in fact describes a ``density of states'' that is not positive definite.  
A variety of alternatives have been considered---perhaps the sum should include  point particle 
configurations, for instance \cite{Fjelstad,Benjamin}, or perhaps one should average over boundary 
conformal field theories \cite{Cotler,Chandra}---but there is  not yet a conclusive result.

Although the area is far less developed, similar questions have been explored for $\Lambda=0$.
Here, the asymptotic algebra is the BMS${}_3$ algebra (\ref{ae8}), which can again be 
interpreted as a symmetry algebra for asymptotic states.  There is a direct analog of the Cardy formula
\cite{Bagchi},
\begin{equation}
\ln\rho(\Delta_L,\Delta_M) \sim 2\pi\Delta_L\sqrt{\frac{c_{\hbox{\tiny\it LM}}}{2\Delta_M}}  ,
\label{c5}
\end{equation}
and cosmological horizon entropies \cite{Bagchi} and the one loop partition function have been computed 
and agree with standard semiclassical expectations \cite{Barnichb}.

The appearance of boundary states in asymptotically flat and anti-de Sitter space has 
also inspired a more general study of gravitational ``edge states'' in regions with finite boundaries.  In
general, when one imposes boundary conditions, one must add boundary terms to the gravitational
action and restrict the symmetries at the boundary.  This again gives rise to ``would-be
gauge'' degrees of freedom, parametrized by transformations that are no longer invariances of the
action.  

For Chern-Simons theory with a compact gauge group, the resulting edge states are well understood: 
they are described by a boundary Wess-Zumino-Novikov-Witten action \cite{EMSS}.  This action
cannot be neglected; it  is needed to ensure that partition functions for two manifolds joined at a 
boundary ``glue'' properly when one sums over intermediate states \cite{Witten_sew}. 

 For (2+1)-dimensional gravity, much less is known, but some preliminary results exist.  Edge modes 
 have been studied in the first order formalism \cite{Geillera,Geillerb}, with boundary modes that include 
both ``would-be diffeomorphisms'' and ``would-be local Lorentz transformations.''  A related 
piecewise flat discretization of space also incorporates edge modes, which are again central 
for ``gluing'' adjacent cells \cite{Freidel}.  In a rather different approach, a reduced phase space ADM 
quantization of the (2+1)-dimensional causal diamond has been carried out \cite{Andrade}, and
although the Hamiltonian dynamics is not yet understood, the BMS${}_3$ group again makes an 
appearance.   And in yet another approach, discrete Regge calculus calculations have led to the 
emergence  of edge modes \cite{Riello} and a Liouville-like boundary theory \cite{Bonzom}.   
 
\section{Conclusion}

Perhaps the most important lesson from (2+1)-dimensional quantum gravity is that it exists.  This 
may not always be the case; the difficulty of finding a boundary field theory for asymptotically 
AdS space has led some to speculate that there is no purely gravitational quantum theory in that 
setting.  At least for spatially compact manifolds, though, there are clearly sensible answers.  
While these are physically very different from what we expect in 3+1 dimensions, they demonstrate 
that the notorious conceptual problems surrounding quantum gravity are not insurmountable.

A second lesson is that (2+1)-dimensional quantum gravity is not unique.  Path integral formulations
allow topology change; canonical quantization does not.  Within canonical quantization, the
loop representation differs drastically from Chern-Simons quantization.  Even within reduced
phase space ADM quantization, different operator orderings lead to Hamiltonians with different 
transformation properties and different spectra.

The (2+1)-dimensional theory has offered interesting perspectives on a number of broader issues. 
The explicit implementation of covariant canonical quantization for the torus universe suggests
a new way to consistently construct time-dependent observables for arbitrary slicings.  The
asymptotically AdS case introduced the key role of gravitational edge states.  Work on the 
Wheeler-DeWitt equation has shown that spatial nonlocality is a far more serious problem than had 
been appreciated.   Sums over topologies have hinted that the divergences due to the growth in the 
number of topologies are important, but perhaps in somewhat unexpected ways \cite{CarlipHH}.  

Finally, we are left with a plethora of interesting questions.  The boundary theories for asymptotically
AdS and asymptotically flat spacetimes remain mysterious, and work on edge modes for finite
regions has just begun.  There are interesting classical descriptions of big bang singularities
\cite{BenGuag,Mondal}, but no understanding of whether they are ``resolved'' by quantum gravity.
A more systematic summation over topologies might tell us much more about topology change,
and perhaps about the cosmological constant.  There is some evidence that lengths may be quantized
\cite{FLR,Achour,DupuisG,Wielandx}, but the results are mixed, and a deeper understanding is
needed.  And, of course, technical problems remain in each particular approach: how to calculate
or approximate the higher genus Hamiltonian in ADM quantization, for instance, or how to gauge
fix the Wheeler-DeWitt inner product.

\begin{flushleft}
\large\bf Acknowledgments
\end{flushleft}

 This work was supported in part by Department of Energy grant DE-FG02-91ER40674.

\end{document}